\def\year{2022}\relax
\documentclass[letterpaper]{article} 
\usepackage{aaai22}  
\usepackage{times}  
\usepackage{helvet}  
\usepackage{courier}  
\usepackage[hyphens]{url}  
\usepackage{graphicx} 
\usepackage{natbib}  
\usepackage{caption} 
\DeclareCaptionStyle{ruled}{labelfont=normalfont,labelsep=colon,strut=off} 
\usepackage{algorithm}
\usepackage{algorithmic}

\usepackage{newfloat}
\usepackage{listings}
\floatstyle{ruled}
\newfloat{listing}{tb}{lst}{}
\floatname{listing}{Listing}

\usepackage{xspace}
\usepackage[most]{tcolorbox}

\title{Temporal Network Analysis of Email Communication Patterns\\ in a Long Standing Hierarchy}
\author{
    Matthew Russell Barnes\textsuperscript{\rm 1},
    Mladen Karan\textsuperscript{\rm 1},
    Stephen McQuistin\textsuperscript{\rm 2},
    Colin Perkins\textsuperscript{\rm 3},
    Gareth Tyson\textsuperscript{\rm 4},
    Matthew Purver\textsuperscript{\rm 1,5},
    Ignacio Castro\textsuperscript{\rm 1},
    Richard G. Clegg\textsuperscript{\rm 1}
}
\affiliations{
    \textsuperscript{\rm 1}Queen Mary University of London, London, UK\\
    \textsuperscript{\rm 2}University of St Andrews, Fife, UK\\
    \textsuperscript{\rm 3}University of Glasgow, Glasgow, UK\\
    \textsuperscript{\rm 4}Hong Kong University of Science and Technology (GZ)\\
    \textsuperscript{\rm 5}Jožef Stefan Institute, Ljubljana, Slovenia.\\
}

\begin{document}

\maketitle

\begin{abstract}

An important concept in organisational behaviour is how \emph{hierarchy} affects the \emph{voice} of individuals, whereby members of a given organisation exhibit differing power relations based on their hierarchical position.
Although there have been prior studies of the relationship between hierarchy and voice, they tend to focus on more qualitative small-scale methods and do not account for structural aspects of the organisation.
This paper develops large-scale computational techniques utilising temporal network analysis to measure the effect that organisational hierarchy has on communication patterns throughout an organisation, focusing on the structure of pairwise interactions between individuals. 
To this end, we focus on one major organisation as a case study --- the Internet Engineering Task Force (IETF) --- a major technical standards development organisation for the Internet.
A particularly useful feature of the IETF is a transparent hierarchy, where participants take on explicit roles (e.g., Area Directors, Working Group Chairs), and because its processes are open we have visibility into the communication of people at different hierarchy levels over a long time period.
Exploiting this, we utilise a temporal network dataset of 989,911 email interactions among 23,741 participants to study how hierarchy impacts communication patterns.
We show that the middle levels of the IETF are growing in terms of their dominance in communications. 
Higher levels consistently experience a higher proportion of incoming communication than lower levels, with higher levels initiating more communications too. We find that, overall, communication tends to flow ``up" the hierarchy more than ``down". 
Finally, we find that communication with higher-levels is associated with future communication more than for lower-levels, which we interpret as ``facilitation".
We conclude by discussing the implications this has on patterns within the wider IETF and the impact our analysis can have for other organisations.
\end{abstract}

\section{Introduction}
\noindent 
The nature of communication within an organisation is indicative of the working practices that exist between participants. For basic functionality of the organisation, there must exist useful communication amongst people who are working towards shared goals. In an organisation with a formal hierarchy, where some participants have official roles, the position of people within the hierarchy may impact communication between them.  

In the field of Organisational Behaviour \citep[OB,][]{pfrombeck2023hierarchy} there is a concept known as the ``voice", which is conceptualised as the ability for an employee of an organisation to ``speak up", i.e give their opinion. There is an area of OB research which has a keen interest in how hierarchy affects the ``voice" of individuals. They draw some conclusions about the effect of organisational structure on the voice of employees, and this paper builds on these conclusions with new analysis techniques.

The aim of this paper is to quantitatively determine the effect that organisational hierarchy levels have on communication patterns throughout an organisation. We model communication as a network where a node represents an individual and a directed edge represents a communication sent from one individual to another~\cite{panzarasa2009patterns,viswanath2009evolution,klimt2004enron}. 
By repeating the analysis across time, we can build a full temporal network~\cite{masuda2016guide} of communication interactions between people in the system.In this study we consider only the network structure and the status at that time of the individuals, not the content of messages.


The type of hierarchy we are interested in has participants with stratified roles for organisational purposes. Therefore, this paper uses as its case study interactions between participants of the Internet Engineering Task Force (IETF), an organisation that develops technical standards for the operation of the Internet. The IETF operates primarily using a number of email lists, plus in-person plenary meetings three times a year. We study how the email communication is impacted by the hierarchical roles of individuals. According to the IETF mission statement~\cite{rfc3935}, the IETF is made up of volunteers who collaborate to develop consensus-based technical standards. This collaboration should be evident in their communication practices.
The nature of the IETF mailing list dataset~\cite{khare2022web} enables unique insights as it is both long-standing (a mostly complete record of all emails has been kept since 1980), has been combined with comprehensive metadata (the hierarchical status of participants for example), and is high impact (the IETF creates critical infrastructure standards relating to the technical operation and design of the Internet).

To the best of our knowledge, ours is the first paper that uses graph structure to investigate the effects of hierarchy on individuals and their communication patterns. Even organisational structural analysis of how communication is affected by hierarchy has rarely been performed in the OB literature. However, we have put together important hypotheses from OB into the research questions, RQ1-RQ4, in this paper. 

\section{Research Questions}
The first hypothesis focuses on the impact of the ``steepness"~\cite{anderson2010functions} of the hierarchy. This refers to the fewer people in higher hierarchy levels the more steep the hierarchy is, i.e a centralised control of decision making in higher levels. The opposite of this is a ``diffused"~\cite{hussain2019voice} hierarchy where the large amount of people in higher levels stifle lower level voice by creating ``bystanders" who find it hard to speak up. In general, the conclusions about steep hierarchies are confused, where some papers conclude a positive and some a negative effect of steep hierarchies. We explore what this means for the IETF in RQ1.

The second hypothesis regards the ``power distance"~\cite{li2021does,duan2018authoritarian,guo2020inclusive} of individuals, which is the number of hierarchy levels between people who are communicating. The focus is on the impact this has on people's voice or lack thereof when the distance is large versus small. The conclusions from this research area usually suggest that even small power distances have a large effect of reducing the voice of the lower of the two levels. We tackle this for the IETF in three different ways outlined in RQ2-4.

The IETF being a voluntary organisation gives a new perspective on these hypotheses, as the existing literature focuses on commercial organisations. In contrast to the above conclusions, we demonstrate that despite the hierarchy becoming more ``diffused", participants in the higher levels in the IETF hierarchy perform a ``facilitator" role, promoting discussion with those they interact with and receiving a benefit themselves as a result. We also show that higher level participants tend to be a focus of discussion with remarks directed ``upward" toward them whereas Regular Participants (RPs) engage more in group-discussion with less focus on an individual. This suggests that the IETF does not conform to the notion that power distance has a negative effect on communication, nor that a more diffused hierarchy reduces the voice of lower levels.

\begin{itemize}
\item \textbf{RQ1: Are higher levels of the hierarchy growing or becoming more centralised? Are they associated with an increased domination of the conversation?}

The effect the steepness of an organisation's hierarchy has on the communication within is not well understood. This RQ is important because once we know whether the IETF hierarchy is becoming steeper, we can look at the effects this is having on the communication patterns of participants.

In this RQ, we look at the number of individuals and/or roles that inhabit each hierarchy level. We also consider whether individuals in higher levels are more active in the conversation and quantify the evolution of the mailing list activity within each level of the IETF hierarchy. We find that the middle level, Working Group Chairs (WGCs), is gaining in overall proportion of activity, whilst the lowest level, RPs, is decreasing. The top level, Area Directors (ADs), remains constant in proportion. The increase in WGC activity coincides with an increase of WGCs per Working Groups (WGs).

\item \textbf{RQ2: What is the association between the organisation hierarchy and general communication patterns?}

The communication patterns that different levels in the hierarchy experience may be  indicative of the cooperation between levels~\cite{li2021does,duan2018authoritarian}. For instance, if participants at higher levels send lots of email while receiving little, this suggests a lack of receptiveness of higher levels to the suggestions of lower levels, and low confidence of lower levels to voice their opinions. In contrast, we hypothesise that collaboration between layers is high, as outlined in the IETF mission statement, which will look more like the opposite of this example.

In this vein, we compare the ways that different roles communicate, looking at the tendency for communications to be purely inbound, outbound or use more complex patterns. We count three edge motifs for different hierarchy levels as described in section~\ref{sec:motif}. We find that higher levels consistently experience a higher proportion of incoming communication than RPs, and all levels send outward communication at similar proportions over time. Also, the proportion of people who begin email threads are split by hierarchy level. Higher levels show a disproportionate tendency to originate email threads. 

\item \textbf{RQ3: How do people with differing roles communicate with each other? Does information flow ``up" or ``down" the hierarchy?} 
Our third hypothesis is that the cooperative nature of the IETF will be evident in the effect that power distance~\cite{li2021does} has on communication throughout the organisation. The raw proportion of the communication between hierarchy levels helps us to understand the confidence of lower levels to exercise their voice. This proportion, for a cooperative and voluntary organisation should be higher for lower hierarchy levels, showing a confidence in their voice.

We categorise edges based on the hierarchy levels they originate from and are sent to. Ratios are then calculated of ``upward" versus ``downward" communication between RPs, WGCs and ADs. We find that the lower the level, the more of a skew to upward communication. Communication tends to flow ``up'' the hierarchy more than ``down''. 

\item \textbf{RQ4: What is the impact that individuals have on their direct contacts' activity over time? How does this vary across hierarchy level?} 

An important question~\cite{li2021does,guo2020inclusive} for any organisation is whether higher level participants encourage those at lower levels in the hierarchy to properly engage. The IETF is cooperative and voluntary by design, therefore we hypothesise a high level of encouragement from higher levels to lower. We measure whether activity in one time period is associated with activity in a subsequent time period both for individuals and for neighbourhoods. If those who communicate with higher levels receive a boost in their communication as a result, this is an indication of such encouragement.

In particular, we perform Mobility Taxonomy analysis to determine the effect individual IETF participants have on their future number of connections, and the average number of connections of their neighbours. We find that WGCs have an increased tendency, versus RPs, to remain active (\textit{Mobility}), an increased tendency for their neighbours to be active subsequent to high WGC activity (\textit{Philanthropy}) and for them to gain activity after their neighbours are active (\textit{Community}).
We interpret this as WGCs filling a ``facilitator" role on the mailing lists.

\end{itemize}

\section{Related Work}
\subsubsection{Temporal Networks}
Analysing temporal networks is an emerging field of research~\cite{masuda2016guide} which allows for networks to evolve in time, bringing them more in line with natural networks which do not remain static. For instance, some have analysed the time evolution of important nodes in a network~\cite{fire2020rise}, whether their neighbours are similarly important~\cite{pedreschi2022temporal} and some others the correlations of the trajectory of snapshots of networks over time~\cite{lacasa2022correlations}. In this paper we bring new analysis techniques, that expand on the idea of tracking the evolution of node importance, to an email list corpus.

\subsubsection{Temporal Motifs}
A form of temporal network analysis that is key to understanding communication in social networks is computing temporal motifs~\cite{paranjape2017motifs,kovanen2011temporal} between nodes. This technique involves look at the sequence in which edges attach to nodes over time which lets us see how conversation happens with fine resolution. Applications include understanding the diffusion of information through a social network using undirected motifs~\cite{sarkar2019using}, and analysing the collaboration and scientific mobility of co-authorship networks~\cite{boekhout2021investigating}. To the best of our knowledge, temporal motifs have not yet been used for analysis of communication in social networks which means knowledge of high time-resolution direct communication between individuals is ripe for the picking.

\subsubsection{Communication Networks}
Building social interaction networks out of email communication data is a well trodden field~\cite{belanger1999communication, mcpherson2001birds}, and the introduction of time evolution into the analysis~\cite{panzarasa2009patterns,viswanath2009evolution} is recent but has quickly gained interest. However, the analysis of email based communication is less prevalent with the notable exception of the well-known, and oft analysed, Enron emails~\cite{klimt2004enron}. The communication dataset we examine is interesting for several reasons: it is a long duration email dataset, it is annotated with meta data representing participant roles, and it represents discussion in a decentralised consensus driven organisation~\cite{rfc7282,rfc3935} rather than a business with a traditional management hierarchy.

\subsubsection{Voice}
The effects of hierarchy on the ``voice" of individuals within an organisation is a key aspect~\cite{pfrombeck2023hierarchy} of Organisational Behaviour (OB) research. A steep~\cite{hussain2019voice} hierarchy can have a negative effect on voice by reducing the variety of perspectives~\cite{anderson2010functions} in higher levels, or positive effect where decisions can be made quickly. However, a diffused hierarchy may also have negative effects~\cite{hussain2019voice} on voice by contributing to lower level ``bystanders" who let the higher levels do the talking, or a positive effect with a variety of approaches to management. Similarly, a large ``power distance" can cause problems  to lower level voice if the higher levels are ``authoritarian"~\cite{duan2018authoritarian}, however if they are ``benevolent"~\cite{li2021does,guo2020inclusive} then power distance is less of a problem. We aim to test these ideas in regard to the voluntary and collaborative nature of the IETF organisation, whereas the literature mostly focuses on traditional organisations.

\subsubsection{IETF}
The IETF has made public much useful data for analysing their organisation. In this paper we focus on the mailing list social interaction graph and hierarchy data, but there is also data readily available about the RFC Series\footnote{\url{https://www.rfc-editor.org/} -- The term RFC used to stand for Request for Comments, but over the years the series has become a publication venue for completed Internet standards and similar documents, not a discussion forum, so RFC is no longer expanded.} where the IETF publishes technical standards for the Internet. Multitudes of public metadata are also available and have been analysed in some recent publications. First, \cite{mcquistin2021characterising} concentrates on predicting the success of RFCs based on their complexity and the time it takes to finish them. Second, \cite{khare2022web} analyses the effect of influence on the adoption of an RFC by a WG. Then~\cite{khare2023tracing} focuses on predicting the influence of individuals based on linguistic analysis of the content of their emails. Finally~\cite{karan2023leda} characterises the mailing list data as dialog acts for the study of decision-making mechanisms. This paper by contrast is the first to use these temporal graph analysis techniques on the dataset and the first to look at the interaction between the temporal graph and the hierarchical status of its participants.

\section{Methodology}
\label{sec:method}
\subsection{Temporal Graphs}
Our analysis is grounded in temporal graph analysis techniques, which model how interactions (edges) between entities (nodes) evolve over time.
A temporal graph $G$ is defined as existing from time $0$ until $\infty$, with nodes $V$ and temporal edges $E$. It is built out of edge events $(n,m,t)$ where $n$ and $m$ are nodes which are connected by an edge at time $t$. The edges are directed and are defined by a pair of nodes $e=(n,m) \in E$, where the ordering of the nodes is important and $n,m \in V$ but $n\neq m$. If we let $0\leq t_1\leq t_2\leq\ldots\leq t_n< \infty$, where $t_n$ is the number of events, then the edge events make up a set
$$T = \{(n_i, m_i, t_i): i=1,2,\ldots\}$$
where the same edge $(n, m)$ can exist in many events, and at multiple times. This set does not need to be ordered by index, as what is important is the times they occur.

The events can be aggregated by time, i.e. all of the events which occur at time $t$, into a graph $G(t) = (V(t),E(t))$ containing only the edges which appear in the appropriate events. This is not necessarily a ``complex" graph, as it could be the case that only one edge exists at a particular $t$. These snapshot graphs can be used to represent the temporal graph
$$G(0,\infty) = \{G(t_1),G(t_2),\ldots)\}$$
and in a similar way, we define the graph $G(q,r) = (V(q,r),E(q,r))$ that exists in ``time window'' $(q,r)$, from time $q$ to $r$, where $0<q<r<\infty$, as the aggregation of all edge events $(n_i, m_i, t_i)$ where\footnote{Excluding the timestep $r$ removes double counting of adjoining time windows.} $q\leq t_i < r$. The number of edges that node $n$ in $V(q,r)$ takes part in is the degree of that node $k_n(q,r)$.

\subsection{Activity}
The activity of a node in a temporal graph is closely aligned with the definition outlined above. Knowledge of the activity of nodes over time allows us to draw conclusions about the level of participation in a network.
A node is considered active if it has appeared in at least one edge
in the time window $V(q,r)$. The level of activity of a node is defined as how many edges, including duplicates, that they participate in during $V(q,r)$. This is different to the degree within the time window as duplicates are not considered.

\subsection{Mobility Taxonomy}
\label{subsec:mob_tax}
We aim to measure the association of the activity of nodes, and their neighbours, with their activity in a subsequent time period. To determine this association we use the temporal graph analysis technique called the Mobility Taxonomy~\cite{barnes2023measuring}, which involves correlating the degree and average neighbourhood degree (ND) between two adjoining time windows, both half of the chosen time window length for other analysis. The temporal graph analysis tool Raphtory~\cite{steer2020raphtory} is used to calculate the raw degree and ND numbers before performing these correlations. 

There are six combinations to correlate node degree and ND in time window one and two, the combinations are outlined in detail in~\cite{barnes2023measuring}. The four measures used in this paper are \textit{Mobility}, \textit{Neighbour Mobility}, \textit{Philanthropy} and \textit{Community}. 
Each represents the correlation of the degree of a node (or set of nodes) at an earlier or later period of time. \textit{Mobility} can be thought of as the tendency for a node that is active in the first time window to be equally as active in the second. \textit{Neighbour Mobility} is similar but for a node's neighbourhood. \textit{Philanthropy} can be thought of as the tendency of a node's neighbourhood to be active in window two if that node is active in window one. Finally, \textit{Community} is the reverse of \textit{Philanthropy} as the activity of a node in window two is compared to that node's neighbours activity in window one.

Only the nodes which are active in the first window are considered for the second window. Nodes inactive in the second time window are said to have a degree of 0. The ND in the second time window is taken over the same set of neighbours that existed in the first; that is we look at a node's neighbours in the first window and measure their degree in the second for consistency of comparison. 

\subsection{Temporal Motifs}
\label{sec:motif}
Looking at the sequence in which edges attach to nodes over time lets us see how conversation happens with fine resolution. Temporal Motifs are sub-graphs where all edges occur within a certain time.  
In this paper, we analyse every possible combination of three temporally ordered links~\cite{paranjape2017motifs} between at most three nodes.
The time order in which the nodes occur is important. 
The direction of edge 1 is arbitrary, but every edge temporally after it has their direction oriented with respect to it. 

\begin{figure}
    \centering
    \includegraphics[width=0.95\columnwidth]{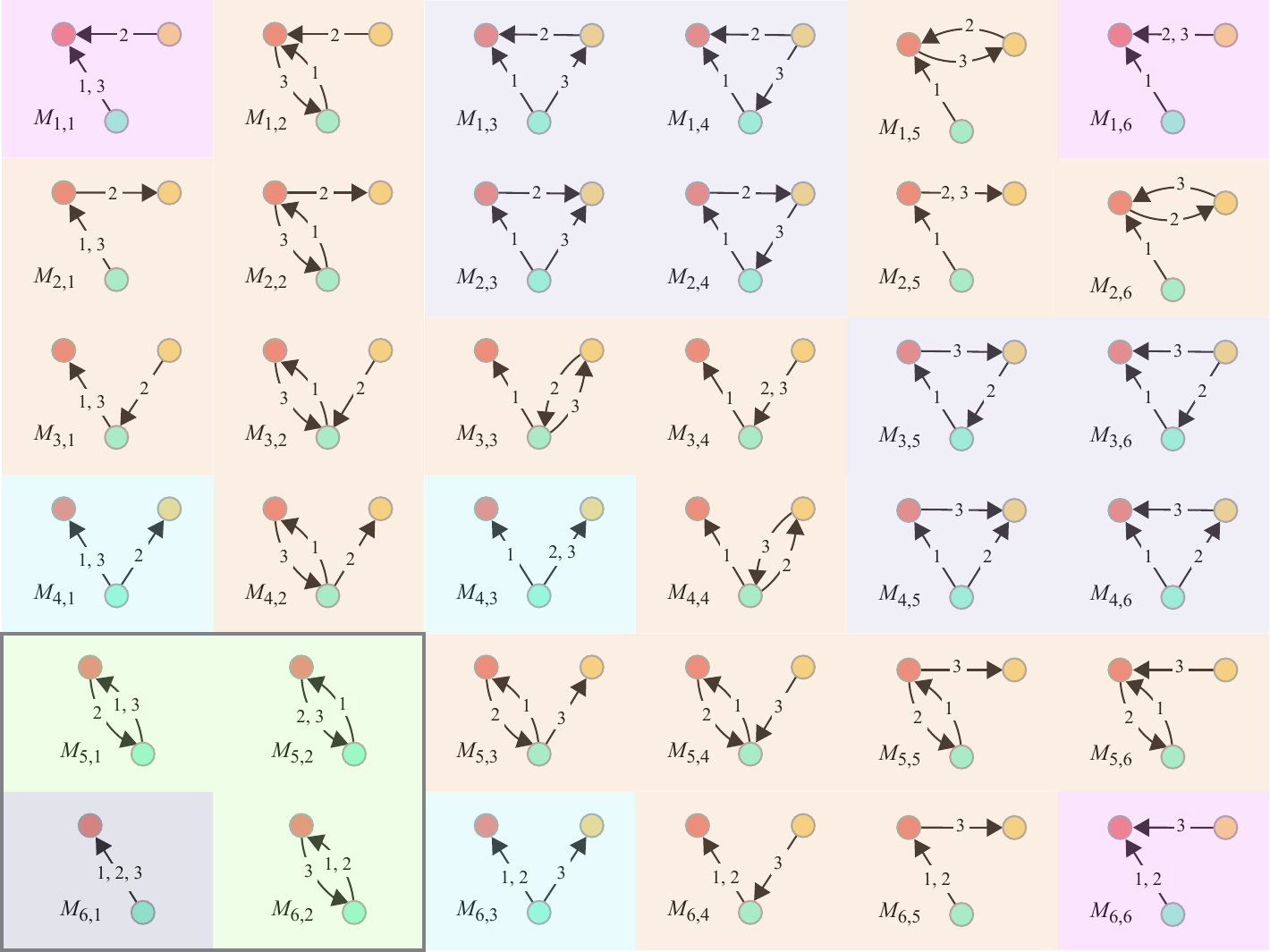}
    \caption{Classification of three edge motif types with the numbers representing the order of communication, and $M_{i,j}$ simply refers to the motifs positional square in this plot. For example $M_{1,6}$ represents the cyan node emailing the red followed by the yellow emailing the red twice. 
    Light blue squares are  ``Outward Star" motifs; purple are ``Inward Star" motifs; orange are ``Mixed Star" motifs and grey are ``Triangle" motifs. The box in the bottom left hand corner represents motifs with only two nodes.}
    \label{fig:motif_types}
\end{figure}

Figure \ref{fig:motif_types} shows all possible types of interaction and classifies them into five types: motifs with two nodes; ``Outward Star", announcements or dissemination of information; ``Inward Star", questions or condensing of information; ``Mixed Star", one-on-one discussion; and ``Triangle" is group discussion with no individual as the focus. Motifs with only two nodes  are rare in this network and we do not present the results.

Raphtory~\cite{steer2020raphtory} is used to count the prevalence of these motifs. We start with the list of 36 different combinations of three directed edges shown in figure~\ref{fig:motif_types}. The graph is then split into time windows and then the motifs are counted combinatorially for each window; each of a node's edges are followed from their origin to three edge steps away. Finally, we combine the motif counts using the categories shown in figure~\ref{fig:motif_types} by the colour of the squares.

\subsection{Social Interaction Graph}
The mailing list dataset for our case study~\cite{khare2022web} characterises the 989,911 interactions among 23,741 IETF participants in 861 email lists from 04/01/1980 until 17/04/2021, and was gathered from the IETF Mail Archive which is publicly accessible on the web\footnote{\url{https://mailarchive.ietf.org/} -- see Section \ref{sec:ethics} for discussion of the ethics of data access.} and in machine readable form via IMAP. The email archive contains public communications only. Note that there may be relevant private discussions not considered here. The data file contains an identifier for both the sender (\texttt{From:}) and receiver (\texttt{To:} and \texttt{Cc:}) of emails, the identifiers for each message sent, the timestamp of the interaction, and the mailing list each email was sent to. 

\begin{figure}[t]
\centering
\includegraphics[width=0.8\columnwidth]{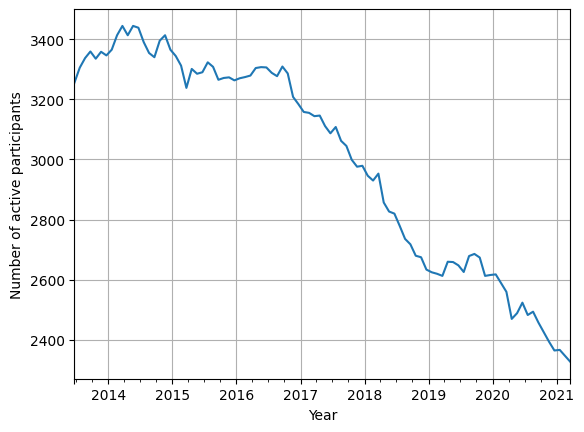}
\caption{Number of active participants in the IETF with active defined as having sent or received an email within the previous year.}
\label{fig:activity_IETF}
\end{figure}

There exist upwards of 250 concurrent mailing lists in the IETF which consist of approximately 2500--3500 active participants at a time, see Figure~\ref{fig:activity_IETF}. The participants are spread amongst around 180 WGs, see Section~\ref{subsec:rq1}, and about two thirds of the mailing lists are specific to WGs while others are more general or organisation-wide. 

We leverage this data to build a temporal social interaction graph of replies to emails in the mailing list where IETF participants are nodes and email replies are the edges. Note that we ignore the first email of each thread if the receiver is the mailing list itself rather than a specific IETF participant.\footnote{Data pertaining to the originator of email threads is retrieved for Figure~\ref{fig:origin_emails}.} As the minimum time resolution of the social interaction graph data is one day this does mean the accuracy of the measurement of activity will decrease as the time windows expand in length. Our chosen window length is one year which slides forward in time by one month. The year was picked as a time resolution because it has a reasonable ``smoothing" effect but still shows good time resolution in the data. The data analysed over a month long time window told a similar story but some graphs were much ``noisier" because less data was available.

\begin{figure*}[t]
\centering
\includegraphics[width=0.95\textwidth]{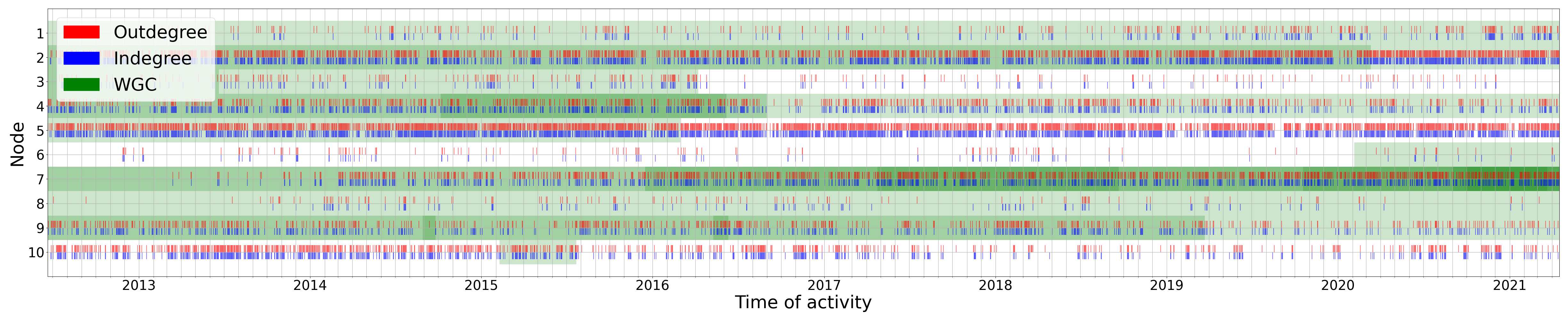}
\caption{Activity for the ten participants of the IETF who are active longest and become WGCs: the blue ticks (in degree) represent emails received as replies and the red ticks (out degree) represent emails sent as responses to others. Green indicates a period that an individual is a WGC with a darker green indicating WGC for multiple WGs.}
\label{fig:tickgraph}
\end{figure*}

An insight into the richness and diversity of the data is given by Figure~\ref{fig:tickgraph} that shows email reply activity for ten individuals. The ten individuals are those who have been recorded as being a WGC with the earliest start dates in the dataset. Some individuals are highly active throughout the whole period considered, for instance nodes 2 and 5, while others have comparatively sparse activity, such as 1 and 3.

\begin{figure}[t]
\centering
\includegraphics[width=0.8\columnwidth]{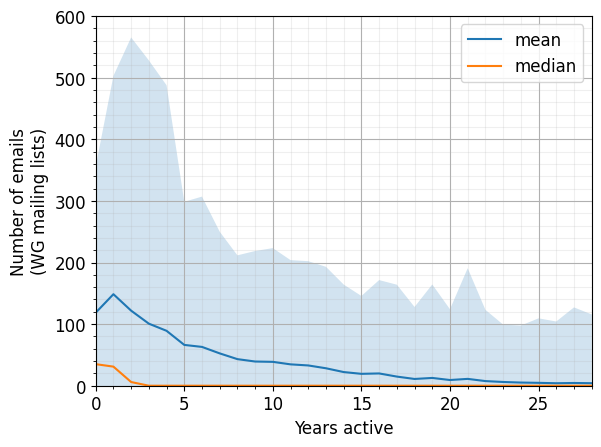}
\caption{The number of replies received to WG mailing lists per year since its inception. The mean/median is taken only over lists that are active for that length of time and the shaded region represents one standard deviation.}
\label{fig:wg_email_activity}
\end{figure}

Figure~\ref{fig:wg_email_activity} gives an insight into the lifecycle of WGs in terms of raw email activity. The large standard deviation is due to, by nature some mailing lists are extremely active while others are relatively quiet. In this plot a ``rise then slowly dwindle" pattern can be seen, where the number of emails within lists has a peak in the mean within the first two years and falls off over subsequent decades.

\subsection{Hierarchy Roles}
\label{sec:roles}
We treat the participants of the IETF as belonging to three roles, from top to bottom: Regular Participants (RPs), Working Group Chairs (WGCs) and Area Directors (ADs). WGCs have an organisational role for one specific Working Group (WG) and ADs have a role organising an Area,\footnote{Some IETF participants take on both positions simultaneously.} which encompasses many different WGs~\cite{rfc2026}. The IETF collects data about when people take on more responsibility in the organisation, which we utilise to assign roles to individuals. Stratifying the network like this allows for comparative analysis between hierarchical levels within the network, and analysing the patterns of communication within and between levels. 

The WGC data is gathered from the IETF Datatracker\footnote{\url{https://datatracker.ietf.org/} – see section \ref{sec:ethics} for discussion of the ethics of data access.} using the ``group events" API for each WG, any WGs where the Datatracker does not contain data about WGCs\footnote{Older WGs may have incomplete WGC data.} are omitted. This data comes in the form of the names of IETF participants becoming WGCs, the names of those who leave the position and timestamps for the change. There are also events for the creation, conclusion and activation of groups, which are used to extend the first and last WGC to the start
and end of the WG's life respectively.

The AD data is gathered from the Internet Engineering Steering Group's (IESG) Past Members web page,\footnote{\url{https://www.ietf.org/about/groups/iesg/past-members/} -- the Area Directors collectively comprise the IESG.} and comes in the form of the IETF participant names and timestamps. The timestamps correspond to each IETF meeting, about every 4 months.

Data on the hierarchical roles for WGCs is only available from 21/06/2012.
For consistency, our final dataset covers the period 21/06/2012--7/04/2021, which includes 10,319 distinct IETF participants and a total of 557,236 mailing list interactions.

\subsection{Node Role Classification}
\label{sec:classes}
The timestamped WGC and AD data was used to allocate hierarchical roles for each person's activity. For instance, a role of ``WGC" is applied to a person if the hierarchy data shows that they are \textit{currently} a WGC when they appear in the interaction graph dataset. These roles are first applied directly to the interaction graph, where the minimum resolution of the time data is one day. Although, when the time data is collated into time windows of one year, the roles are only applied if the person is a WGC or AD for the whole window. Any person who has a role for only part of a window is ignored for that window. The Regular Participants (RPs) are distinguished from the other roles by considering only those who \textit{never} become WGCs or ADs in the entire temporal graph. Combining all of these data, a temporal graph with nodes labelled by their current role in the organisation hierarchy is created.

We also label the edges by the roles of nodes at each end. We label each edge as ``up hierarchy" when a node with a lower hierarchical role emails a node with a higher one (e.g. RP emails a WGC or an AD, WGC emails an AD). Conversely, when a node with a higher hierarchical role emails a node with a lower one, we label the edge as ``down hierarchy" (e.g. AD emails a WGC or an RP, WGC emails an RP). Note that communications might also take place off-list, which figure~\ref{fig:activity_IETF} may indicate that is increasingly the case. These are however likely to result in eventual on-list communications as this is where official discussions and decisions take place. To mitigate any potential biases we focus on the ratio of up hierarchy and down hierarchy communications over time rather than absolute numbers. 

\subsection{Working Group Activity}
\label{sec:WG_activity}
We use two methods to estimate the number of WGs over time. For the first method, we re-use the ``group events" data gathered for the WGC roles (see section \ref{sec:roles}) which only contains WGs for which the Datatracker has data on their chairs. Therefore, this method may under count the true number of WGs before WGC data was gathered. 

The other method uses the mailing list dataset to estimate from the activity seen in each WG specific mailing list. A WG mailing list is considered ``active" from when the first email in the interaction graph is sent to it, and is no longer active after the last email. This method may over count as some mailing lists are active even when WGs have not officially started or have officially concluded. Also, the closer the timestamp reaches to the end of the dataset the more likely it is to stop short of its true activity due to future activity not existing in the data. Therefore this method's estimate is truncated at the start of 2021.

Both methods rely on past and future data for their estimates and so large amounts of both will enable greater accuracy. Therefore, there is a lag in accuracy at the start and a drop at the end for both. 

\subsection{Ethical considerations} 
\label{sec:ethics}
Participation in the IETF is subject to agreements and policies that explicitly state that mailing list discussions and Datatracker metadata will be made publicly available.\footnote{See both \url{https://www.ietf.org/about/note-well/} and the IETF privacy policy available at  \url{https://www.ietf.org/privacy-statement/}.}
We use only this publicly available data in our analysis.
We have discussed our work with the IETF leadership to confirm that it fits their acceptable use policies.
We have also made provisions to manage the data securely, and retain it only as necessary for our work.

\section{Results}

\subsection{RQ1: Are higher levels of the hierarchy growing or becoming more centralised? Are they associated with an increased domination of the conversation?}
\label{subsec:rq1}

\begin{figure}[t]
\centering
\small{a)} \\
\includegraphics[width=0.95\columnwidth]{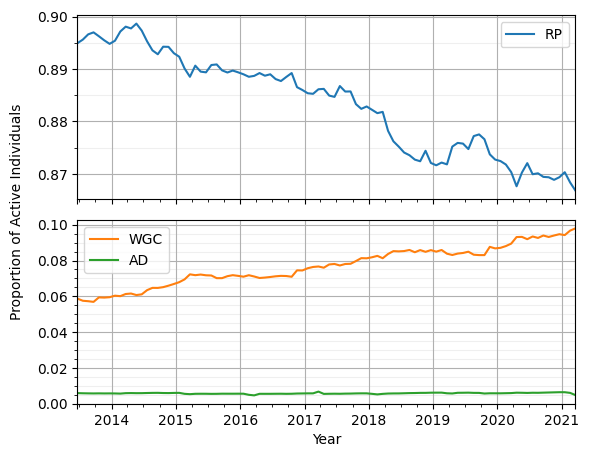} \\
\small{b)} \\
\includegraphics[width=0.95\columnwidth]{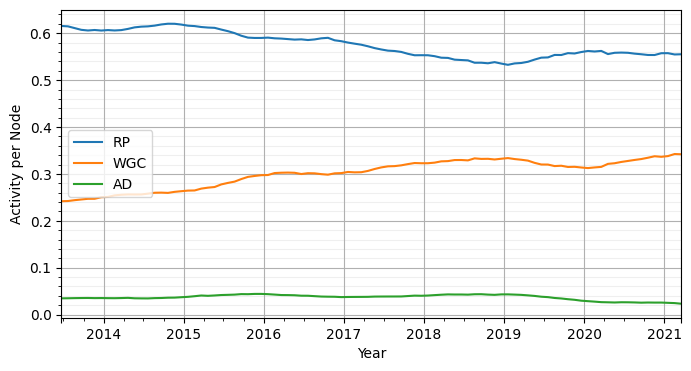}
\caption{Proportions of RPs, WGCs and ADs by a) number of active individuals and b) activity (number of emails sent to mailing lists). 
}
\label{fig:activity_proportions}
\end{figure}

\begin{figure}[t]
\centering
\includegraphics[width=0.95\columnwidth]{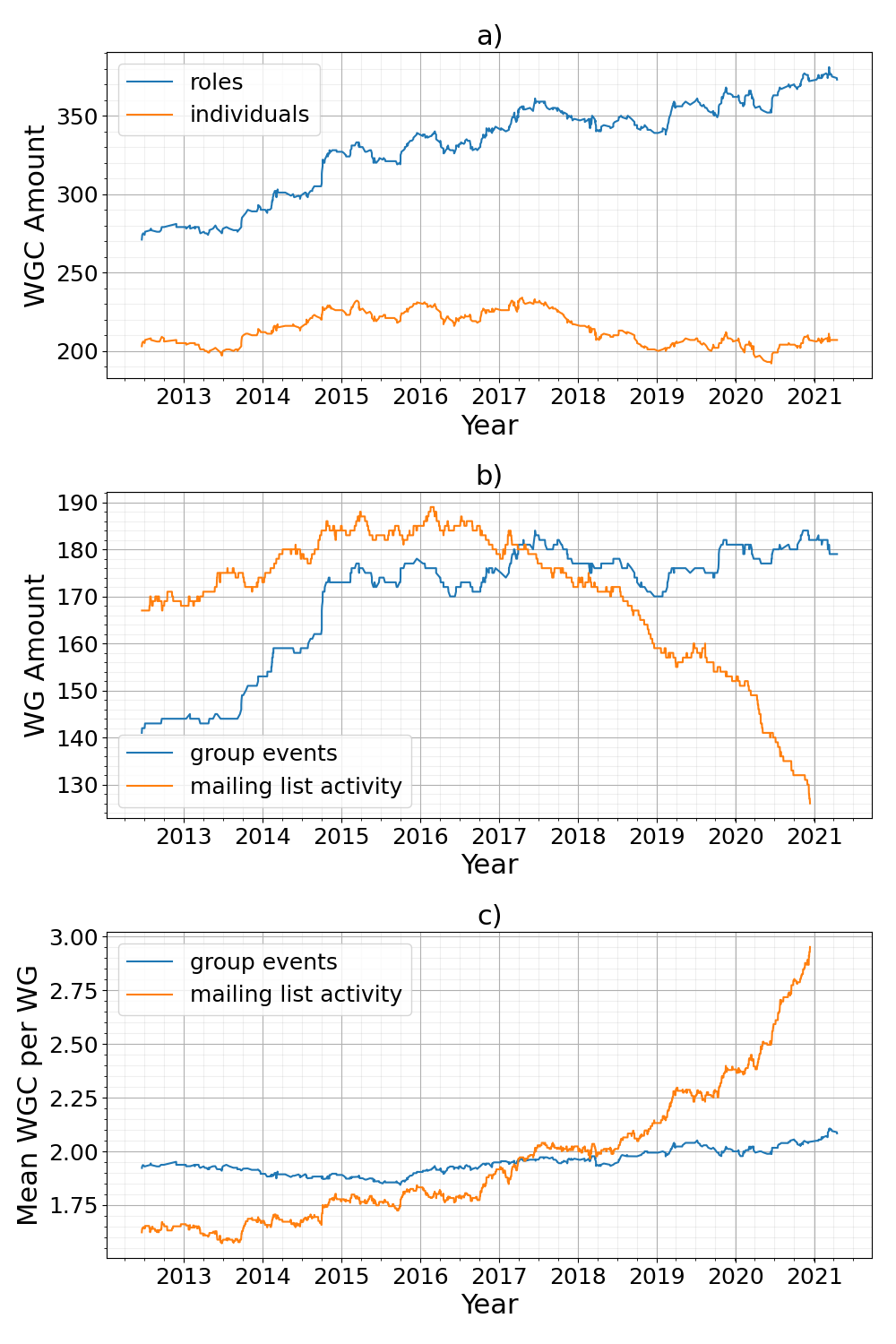}
\caption{Overview of WGC numbers and WG numbers over time. a) shows the number of WGC roles (some individuals hold multiple WGC roles) and the number of individuals who are WGCs b) shows two different estimates of the number of WGs (see section~\ref{sec:method}) and c) shows the mean number of WGCs per WG. Notice the non-zero-based y-axis exaggerates the variation in these figures.}
\label{fig:wgc_over_time}
\end{figure}

For \textbf{RQ1}, to determine the steepness or diffusion of the hierarchy, the proportion of nodes and activity are determined in each hierarchy level. Figure~\ref{fig:activity_proportions}a shows the number of active nodes in each level. The bottom level, RPs, are the largest contingent consisting of 90--86\% of active participants; this proportion is decreasing over time. The next level is WGCs which manage the WGs and are 6--10\% of active participants; this proportion is rising over time. The top level is ADs which manage many WGs within an area and consist of less than 1\% of the total active population of the IETF; this stays constant over time. Therefore, the WGC level is becoming more diffused over time as there are about 15 RPs for every one WGC in the organisation in 2013 versus 8 RPs per WGC in 2021. 

The proportion of activity of nodes in the last year is also plotted and split into hierarchy level in Figure~\ref{fig:activity_proportions}b. It is clear that individuals at higher levels contribute a disproportionate share to communication and this share is increasing over time, which is consistent with findings in \cite{khare2022web}. Therefore, less of a proportion of communication is flowing from the RPs at the same time as the number of WGC roles is increasing. This further suggests that the IETF hierarchy is becoming more diffused, in terms of mailing list communication, which may make cooperation between hierarchy levels more difficult.

Figure~\ref{fig:wgc_over_time}a shows the amount of nodes that inhabit the WGC hierarchy level over our time period. There is a $35\%$ rise in the amount of WGC roles whereas the amount of individuals that fill the WGC roles remains mostly constant.

Figure~\ref{fig:wgc_over_time}b contains two estimates of the number of WGs over time. The estimated number of WGs is likely the most accurate in the middle 50\% of both plotted lines. Their discrepancy is the least (at most 10 WGs) from late 2014 to early 2019. The start of the COVID-19 pandemic may also affect mailing list activity in 2020, and similar trends are visible in the data in \cite{mcquistin2021characterising}. A full explanation on how the data is estimated can be found in section~\ref{sec:WG_activity}.

Figure~\ref{fig:wgc_over_time}c shows how many WGCs exist per WG, using both estimates of the WG amount from Figure~\ref{fig:wgc_over_time}b. It is clear there is a slow a rise in WGCs per WG over time. The ratio rises from between 1.7-1.9 to 2.0-2.2 within the middle 50\%, late 2014 to early 2019.This perhaps represents a growing understanding in the IETF that it is desirable to have two or more chairs per WG in case of illness/unavailability or to avoid conflicts of interest. Therefore, this plot again shows that the hierarchy is becoming more diffused, even if the raw amount of individuals in the WGC roles has remained constant. The multiple roles per individual may result in even more difficulty of RPs to use their voice.

An AD version of Figure~\ref{fig:wgc_over_time} is available, but it is omitted to save space as all three plots stay mostly constant throughout the period. There are about 15 individual ADs split amongst 15 roles in 7-8 areas with about 2 ADs per Area. The set of IETF Areas is broadly fixed, while WGs are created and closed relatively frequently.

\begin{figure*}[t]
\centering
\includegraphics[width=0.95\textwidth]{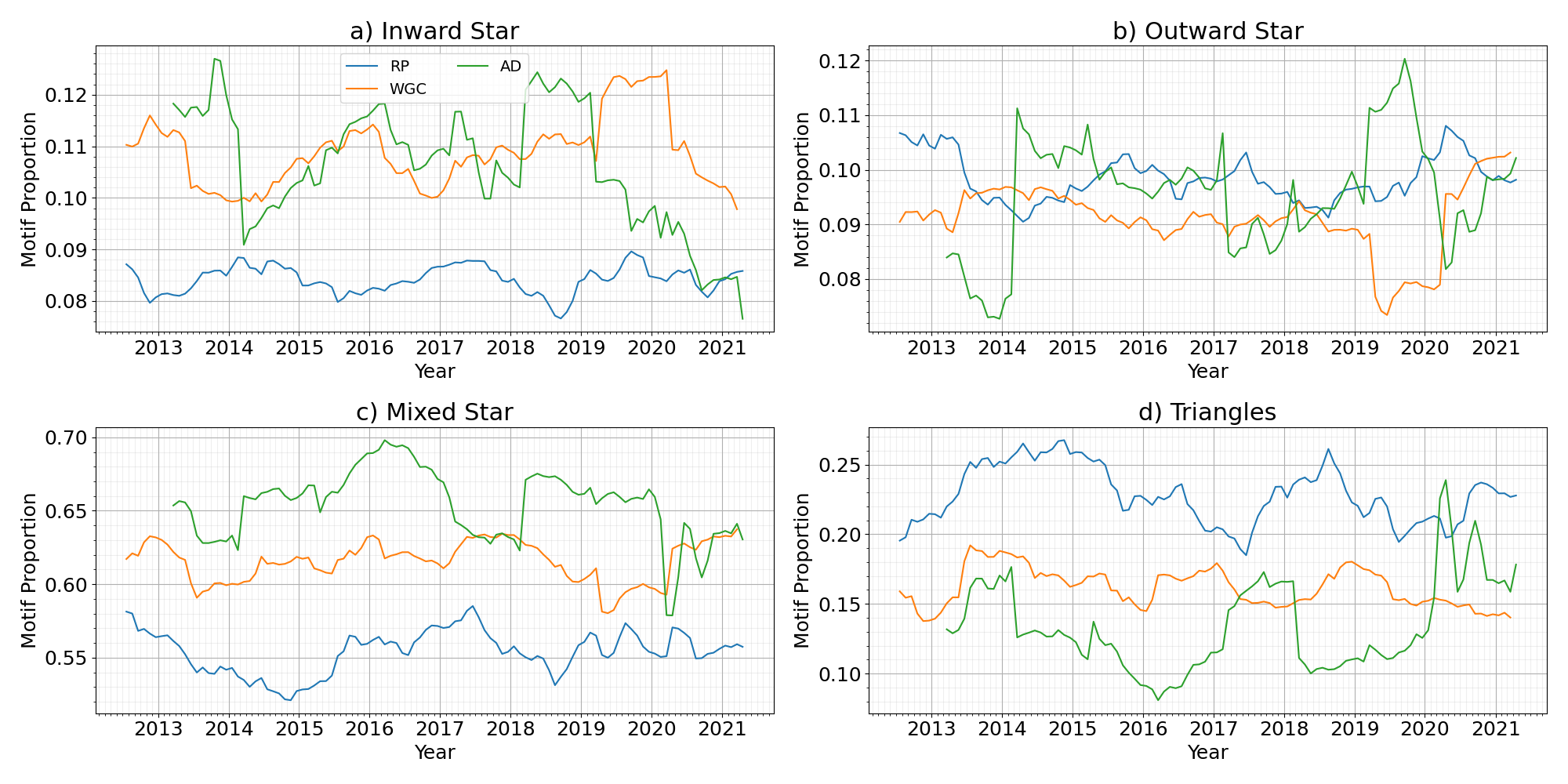}
\caption{Communication patterns analysed by three edge temporal motifs for RPs, WGCs and ADs.}
\label{fig:motif_proportions}
\end{figure*}

\begin{figure}[t]
\centering
\includegraphics[width=0.95\columnwidth]{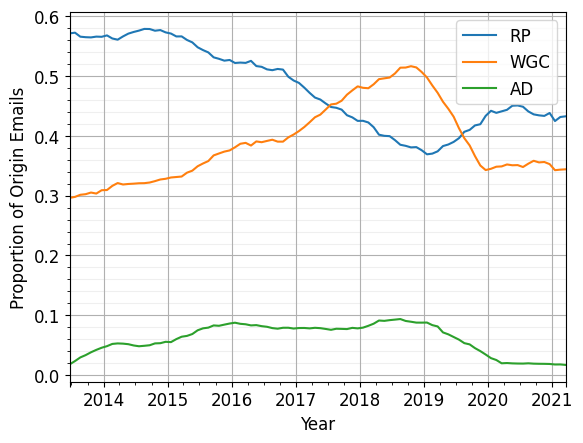}
\caption{Proportion of ``origin" (non reply) emails sent by RPs, WGCs and ADs, this should be interpreted in conjunction with figure~\ref{fig:activity_proportions} showing the proportion of active individuals in each class.}
\label{fig:origin_emails}
\end{figure}

\begin{table}[!tbp]
\centering
    \begin{tabular}{c||c|c}
        Status & Sent & Received \\
         \hline \hline 
        Before WGC & $67.26\pm99.8$ & $70.11\pm103.37$ \\
        \hline
        After WGC & $77.20\pm102.85$ & $85.90\pm107.38$ \\
        \hline
        Before AD & $218.84\pm226.92$ & $230.50\pm214.75$ \\
        \hline
        After AD & $209.21\pm186.46$ & $219.16\pm189.85$ \\
    \end{tabular}
        \caption{Mean number of emails received for one year before/after becoming a WGC (AD) for the first time. The error is one standard deviation.}
    \label{tab:WGC_email}
\end{table}

Table~\ref{tab:WGC_email} shows the emails sent and received for WGC and ADs on mailing lists in the year before they first take that role, and the year after. WGCs and ADs are both active in using the mailing lists, ADs much more so. However, they don't become more active to any major degree after taking the roles. This may suggest that the individuals who take on higher hierarchy level roles are already contributing at a higher level.

In answer to RQ1 the picture is mixed, the number of individuals with WGC roles has remained broadly constant but the number of WGC roles has increased, indicating that individuals willing to become WGC have taken on more such roles. The number of the higher level AD roles has, by design, remained broadly constant over the time studied. The proportion of active individuals with WGC roles has grown from 6\% to 10\% and the proportion of the conversation taken by those roles has grown from 25\% to 35\% over the period studied. Therefore the hierarchy has become more diffused, which may mean RPs experience more difficulty in expressing their voice.

\subsection{RQ2: What is the association between the organisation hierarchy and general communication patterns?}
\label{subsec:rq2}

To quantify the effect of being in each hierarchy level has on network wide communication for \textbf{RQ2}, we calculate the proportion of three node motifs, lasting at most a month, in which each node participates over a year period, see section~\ref{sec:motif}. These motifs are counted, categorised into the three hierarchy levels and then proportions are taken of each motif type. 

Figure~\ref{fig:motif_proportions} shows the proportions of each level for three node ``Inward Star", ``Outward Star", ``Mixed Star", and ``Triangles". The Inward and Mixed Star motifs show about a $4\%$ and $10\%$ increase in proportion for higher levels in the hierarchy versus for RPs. However, about a $10\%$ increase for RPs is seen for Triangles versus WGCs and ADs, and Outward Star proportions remain similar for all levels of the hierarchy.

The larger proportion for WGCs and ADs of Inward and Mixed Star motifs whilst the Outward Star motifs remain similar for all levels suggests that higher levels receive more direct communication. The higher proportion of RP Triangle motifs suggests their discussion is more of a group activity, whereas WGCs and ADs have a larger proportion of one-on-one conversations. Therefore, we interpret the general communication patterns within the IETF as a discussion amongst RPs interspersed with questions sent to and announcements from WGCs and ADs. This indicates some collaboration between layers and confidence of some RPs' to voice their opinions upwards.

\begin{figure}[t]
\centering
\includegraphics[width=0.8\columnwidth]{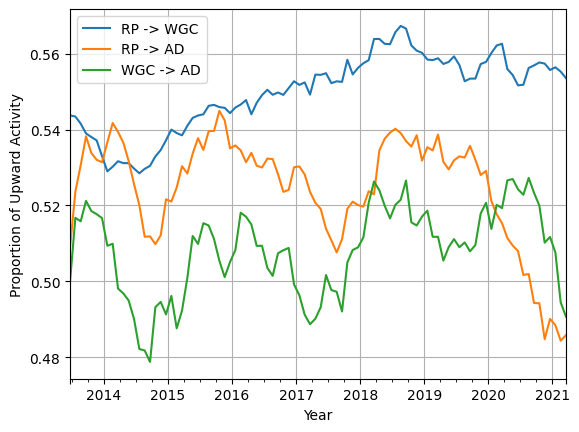}
\caption{Proportion of communications that are ``up" the hierarchy for different groups. A proportion above 0.5 indicates that the majority of communications are ``upward".}
\label{fig:inter_strata}
\end{figure}

\begin{figure*}[t]
\centering
\includegraphics[width=0.8\textwidth]{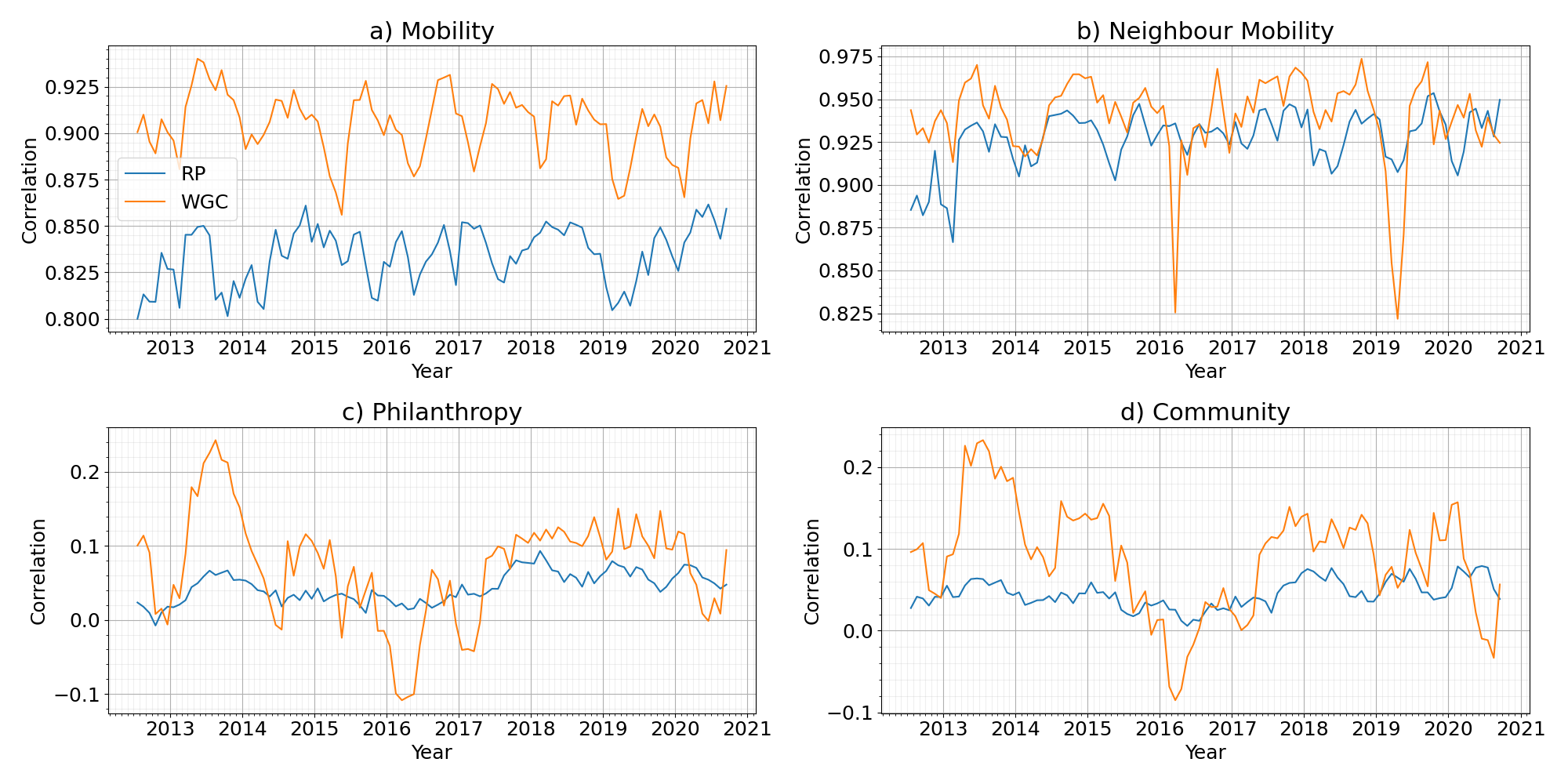}
\caption{Connection between roles and activity for RP and WGC viewed in terms of a) Mobility b) Neighbour Mobility c) Philanthropy and d) Community over time. See section~\ref{subsec:mob_tax} for definitions of these terms.}
\label{fig:mob_tax}
\end{figure*}

Figure~\ref{fig:origin_emails} seems to corroborate this. We calculate the proportion of mailing list threads which nodes in each hierarchy level originate, using a year window, pushed forward by a month each calculation. We see that WGCs and ADs send a disproportionate number of originating emails to the mailing lists. WGCs send 30--50\% whereas they are 6--10\% of active individuals, and ADs 5--10\% versus 0.5\%. 

Therefore, the larger proportion of inward motifs for higher levels is in part due to WGCs and ADs disproportionately originating email threads. Other IETF participants will then reply to the thread, boosting the proportion of inward motifs WGCs and ADs appear in. This may suggest the higher levels are good ``condensers of communication" as they receive more than they send.

In answer to the RQ, combining the interpretations of the two plots, the discussion seems to happen in the following way: The ADs or WGCs are commonly the originators of discussion threads. Where conversation includes a WGC or AD the conversation tends to be directed around them and usually inward towards the WGC or AD. The RPs are more likely to engage in discussions amongst themselves (triangular communication patterns) when higher level individuals are not present. 

\subsection{RQ3: How do people with differing roles communicate with each other? Does information flow ``up" or ``down" the hierarchy?}
\label{subsec:rq4}

For \textbf{RQ3}, we determine the direction communication tends to flow through the hierarchy by looking at inbound and outbound communication flows. The network's edges are categorised based on their source and target node's hierarchy level. For each of the inter-level communication combinations the proportion of communication in the last year going in the ``upward" hierarchy direction is calculated. For instance, if there are six emails from RPs to WGCs and twelve from WGCs to RPs then the proportion upward is $\frac{6}{6+12}=\frac{1}{3}$. 

In Figure~\ref{fig:inter_strata}, these proportions are plotted, and the year long window is again pushed forward by one month to increase the resolution of changes. A proportion of higher than $0.5$ shows more communication flowing up the hierarchy than down. All three levels are mostly upward in direction, with the lowest two levels (RP$\rightarrow$ WGC) having the largest skew in the proportion. The other two lines (RP$\rightarrow$AD and WGC$\rightarrow$AD) show similar periodicity, with the WGC$\rightarrow$AD
line dropping as far below $0.5$ as above showing a large shift in inter-level communication patterns higher up the hierarchy.

The direct measurement of message flow between the hierarchy levels seems to bolster our interpretation that the cooperative design of the IETF is working. This analysis shows that RPs are communicating much more preferentially upwards in the hierarchy, despite the decrease in overall share of RP activity. The increase within the WGC level does not have any noticeable effect on their level of communication, if anything there is a negative relationship. This again suggests both WGCs and ADs perform a kind of ``condenser of communication" or ``facilitator" role. In other words, this analysis suggests the IETF has a ``bottom-up" communication style, where the higher levels encourage the lower levels' voice. 
This is the desired pattern of communication for a volunteer organisation that is looking to grow and encourage participants to engage and gradually take on leadership roles.

Overall for RQ3 then, the pattern observed was communication flowing up the hierarchy and this corresponds to the answer found in RQ2. WGCs and ADs were more likely to receive communication with individuals than send communication to individuals whereas both were more likely than RPs to send out messages to the list in general. 

\subsection{RQ4: What is the impact that individuals have on their direct contacts' activity over time? How does this vary across hierarchy level?} 
\label{subsec:rq3}

Finally, to determine how individuals and their neighbours affect each other's activity levels in subsequent time periods, for \textbf{RQ4}, we take correlations of degree between different time windows using the Mobility Taxonomy~\cite{barnes2023measuring} aspects called \textit{Mobility}, \textit{Neighbour Mobility}, \textit{Philanthropy} and \textit{Community} explained in the previous section. In Figure~\ref{fig:mob_tax}, the time window chosen is one year which is split into two snapshots graphs of six months, the time window then moves forward one month and the process is repeated. The mid point of the year time window is plotted. WGCs and RPs are plotted but the small number of ADs mean that the correlations are too noisy to show meaningful effects. 

The \textit{Mobility} of both RPs and WGCs is plotted in Figure~\ref{fig:mob_tax}a. The high amount of correlation between the degree of an individual in the first half of the time window with the second half shows that all users have a tendency to become more active in subsequent time periods if they are highly active in the previous period, but this is more pronounced in WGCs. This is found to follow a period of about one year for both RPs and WGC, suggesting that a person is likely to be as active as there were a year ago. The \textit{Neighbour Mobility} plot, Figure~\ref{fig:mob_tax}b, shows a similar tendency for neighbourhoods, but with less of a split.

The \textit{Philanthropy} plot, Figure~\ref{fig:mob_tax}c, shows a higher correlation for WGCs than RPs between an individual's degree in the first snapshot with and their ND in the second. This means that individuals who directly interact with WGCs become more active on mailing lists subsequently. The \textit{Community} plot, Figure~\ref{fig:mob_tax}d, shows the correlation between the ND in the first snapshot with the degree in the second. 
The higher correlation for WGCs suggests they benefit from interactions with highly active neighbours by themselves becoming more active in subsequent time periods. These trends for \textit{Philanthropy} and \textit{Community} show a reversal in early 2016 that coincides with a period where the mean number of WGCs per WG and the number of WGC roles was rapidly rising, see figure~\ref{fig:wgc_over_time}. A merging of the Applications area and Real-time Applications \& Infrastructure area in May 2015 may be a cause of this reversal of correlation, as the assignment of WGs to areas was changed which may cause more non-reciprocal WGC communication than normal.

Increased \textit{Mobility} for WGCs suggests that they gain advantage from their role with their activity leading to similar communication activity in the future. Moreover, they experience an increased effect of \textit{Philanthropy} and \textit{Community}. The former suggests that if a WGC is engaged in a higher amount of discussion, then in the future the individuals who they were communicating with will also be highly active. The latter suggests that when WGCs are surrounded by people who are active in discussion, they are likely to be more active themselves in the future too. All of this has less of an effect for RPs who show a lower correlation in terms of all three aspects. Our interpretation of this is that active WGCs are ``facilitating" discussion in the WG mailing lists; both having their activity boosted by their neighbours and, in turn, boosting the activities of their neighbours. This aligns well with the hypothesis that the cooperative and voluntary design of the IETF lends itself well to higher levels encouraging those lower to engage in discussion. 

Overall the answer to RQ4 is that the WGCs have a positive effect on their immediate contacts and those individuals that directly discuss with WGCs are more likely to engage more in discussion in subsequent time periods. Conversely though the WGCs who discuss topics with individuals who are active in discussion are, themselves, more likely to engage more fully in discussion. This points to something like a virtuous cycle of WGCs encouraging and being encouraged by their direct contacts. 

\section{Conclusion}

The goal of this paper was to determine the effect that organisational hierarchy levels have on communication patterns throughout the IETF organisation. We focused on two main hypotheses, first that if the organisational hierarchy is steep, then the communication is centralised causing decision making power to be held by few people, and if the hierarchy is diffused then this stifles the voice of lower levels. Second, we hypothesised that the power distance between people who are communicating has a negative association with the lower level's voice.

In regards to the steepness of the organisational hierarchy, we found that Working Group Chairs (WGCs) are rising as a proportion of active users of the mailing list and in terms of the proportion of total email activity, while the total number of active users is decreasing. Area Directors (ADs) have remained a stable proportion of both active users and email activity over the same period. This shows the hierarchy has become more diffused over the time period, which could mean lower levels find their voice is stifled.

The effect of power distance on communication was investigated in three different research questions. First, we found that WGCs and ADs are characterised by a communication pattern where they disproportionately originate new communication and are more likely to receive replies than send them, whereas Regular Participants (RPs) engage more with triangle motifs. Our second finding is, communication patterns tend to be slightly more ``up" hierarchy than ``down", especially when considering RPs. Together, this suggests that RPs mostly engage more in general group discussion within the IETF mailing lists, and the WGCs and ADs encourage and are receptive to the voice of RPs in one-on-one discussions. Therefore, the distance between hierarchy levels has a small effect on communication between levels as there is a large amount of inter-level discussion.

Finally, using measures like \textit{Philanthropy} and \textit{Community}, we see that active WGCs lead to their neighbours becoming more active in communication and WGCs with active neighbours themselves become equivalently more active, RPs also experience this affect but to minor degree. In early 2016 this association reverses briefly, which may be explained by a merging of two areas. Furthermore, the \textit{Mobility} analysis shows that both WGCs and RPs have a tendency to remain active communicators if they already are such and this is more true of WGCs than RPs. This is interpreted as WGCs embodying a ``facilitator" role within the mailing list discussion.

These three findings suggests that the negative association of power distance is not as pronounced for the IETF as the OB literature finds for traditional commercial organisations. In fact, the facilitation by WGCs of RPs suggests the high levels actively encourage lower level participation in discussion. 

Moreover, combining this with the finding that the organisational hierarchy has a diffused structure, with the potential problems this may cause for lower levels, suggests that collaboration between hierarchy levels is high within the IETF. As the IETF is voluntary and collaborative in its mission statement, this suggests that the their efforts in tackling the problems caused by a diffused hierarchy and power distance have worked well. One suggested improvement for the IETF is to encourage WGCs to get involved in more group discussion in the mailing lists. This may help to boost their level of \textit{Philanthropy}, \textit{Community} and Triangle motifs, which are indicators of healthy discussion. However, we do not advocate for our analysis techniques to be the only metrics maximised for. Moreover, these conclusions come from analysis of only communication structure not content. A future look into content may elucidate differing relationships between hierarchy levels.

This paper shows that important insights can be derived from analysis of communication patterns in organisations through the lens of hierarchies and temporal graphs. This field is a fruitful one and it would be extremely interesting to compare this analysis with other communication datasets where information about individuals' status is available.

\bibliography{hierarchy_icwsm_2023}

\section{Acknowledgements}
The authors wish to acknowledge the support of Moogsoft Ltd. for funding this research.
This work is also supported in part by UK EPSRC under grants SODESTREAM (EP/S036075/1 and EP/S033564/1) and AP4L (EP/W032473/1), and by the Slovenian Research Agency via research core funding for the programme Knowledge Technologies (P2-0103).

\end{document}